\documentstyle[preprint,aps]{revtex}
\begin{document}
\title { Baryon Spin and Magnetic Moments in Relativistic Chiral Quark Models }
\author{H. J. Weber and K. Bodoor }
\address{ Institute of Nuclear and Particle Physics, University of
Virginia, \\Charlottesville, VA 22901, USA}
\maketitle
\begin{abstract}
The spin and flavor fractions of constituent quarks in the baryon octet are 
obtained from their lowest order chiral fluctuations involving Goldstone 
bosons. SU(3) breaking suggested by the mass difference between the strange 
and up, down quarks is included, as are relativistic effects by means of a 
light-cone quark model for the proton. Magnetic moments are analyzed 
and compared with the Karl-Sehgal formulas.    
\end{abstract}
 \par magnetic moments  
\newpage

\section{Introduction }
The nonrelativistic quark model (NQM) explains at least qualitatively many of 
the strong, electromagnetic and weak properties of the nucleon and other 
baryons in terms of three valence quarks whose dynamics is motivated by 
quantum chromodynamics (QCD), the gauge field theory of the strong interaction.
The effective degrees of freedom at low energies are dressed or constituent 
quarks along with Goldstone bosons which are expected to emerge in the 
spontaneous chiral symmetry breakdown of QCD.   
\par
At energy-momentum scales below $\Lambda _{QCD}$ chiral perturbation 
theory~\cite{CPT} allows incorporating systematically the chiral dynamics of 
QCD. Chiral quark models~\cite{WS}, which explicitly include valence quarks as 
effective degrees of freedom and their chiral fluctuations, apply to scales 
from $\Lambda _{QCD}$ up to a (presumed) chiral symmetry breaking scale 
$4\pi f_\pi \approx 1169$ MeV for $f_\pi = 93$ MeV. Other degrees of freedom, 
such as gluons, are integrated out. Clearly, such chiral quark models 
are a drastic truncation of the full gauge field theory with additional 
assumptions about confining potentials. While the NQM description of baryon 
states has not been derived from first principles of QCD, solutions of 
Schwinger-Dyson equations for light quarks with approximations for 
confinement lead to a momentum dependent quark mass $m_q(p^2)$. Such dynamical 
quarks become constituent quarks when $m_q(p^2)$ is approximated by the  
constant $m_q(0)$. The nonrelativistic broken SU(6) spin-flavor symmetry 
of the NQM has been linked with the large $N_c$ limit,~\cite{DJ} where $N_c$ 
is the number of colors. The symmetry structure of the baryon sector of QCD 
is constrained by the condition that pion-baryon scattering amplitudes remain 
finite as $N_c\rightarrow \infty$, and the approximate NQM spin-flavor 
structure of the S-wave baryons with small total spin in the [56] multiplet of 
SU(6) can be understood as a consequence of large $N_c$.          
\par  
Chiral (off-mass-shell, space-like) fluctuations of valence quarks inside 
hadrons, 
$q_{\uparrow,\downarrow}\rightarrow q_{\downarrow,\uparrow}+(q\bar q')_0$, 
into pseudoscalar mesons, $(q\bar q')_0$, of the SU(3) flavor octet of 
$0^-$ Goldstone bosons, were first applied to the spin problem of the proton in 
ref.~\cite{EHQ}. It was shown that chiral dynamics can help one understand not 
only the reduction of the proton spin carried by the valence quarks from 
$\Delta \Sigma =1$ in the NQM to the experimental value of about $1/3$, but 
also the reduction of the axial vector coupling constant $g_A^{(3)}$ from the 
NQM value 5/3 to about 5/4. In addition, the violation of the Gottfried sum 
rule~\cite{Go} which signals an isospin asymmetric quark sea in 
the proton became plausible. The SU(3) symmetric chiral quark model explains 
several spin and sea quark observables of the proton, but not all of them.  
The data~\cite{E143,SMC} call for SU(3) breaking because some of the 
spin fractions such as $\Delta_3/\Delta_8=$5/3 and the weak axial vector 
coupling constant of the nucleon, $g_A^{(3)}={\cal F}+{\cal D}$,  
still disagree with experiments in the SU(3) symmetric case. In~\cite{WSK,SMW} 
the effects of SU(3) breaking were built into chiral quark models and shown to 
lead to a remarkable improvement of the spin and quark sea observables in 
comparison with the data. It was also shown~\cite{WSK} that the $\eta '$ meson 
gives a negligible contribution to the spin fractions of the nucleon not only 
because of its large mass but also due to the small singlet chiral 
coupling constant. We therefore ignore it in the following. In Sect. II we 
describe some of the SU(3) breaking formalism. 
\par
A reduction of the axial charge $g^{(3)}_A$ is known to come from relativistic 
effects as well. Therefore, it is an objective here to include relativistic 
effects along with chiral fluctuations for the proton. Since chiral 
fluctuations in~\cite{EHQ,WSK,SMW} are based on the NQM spin fractions of the 
proton which they improve, they still contain nonrelativistic aspects  
which we replace here by their relativistic analogs obtained from light 
cone quark models. Such relativistic quark models significantly improve 
many predictions of the NQM of which the nucleon weak axial charge is the  
best known example. The relevant formalism is described in Sect. III.  
\par
We also calculate the spin and flavor fractions of several hyperons including 
chiral fluctuations and relativistic effects in Sect.II because from their 
spin fractions one can immediately obtain estimates for their magnetic moments. 
In Sect. IV we compare these magnetic moment values that include chiral 
fluctuations with the Karl-Sehgal formulas involving the corresponding proton 
spin fractions. Such a comparison sheds light on the latter's validity. In 
Sect. V we discuss numerical results more comprehensively. The paper concludes 
with a summary in Sect. VI.
  
\section{ Spin Fractions of Baryons}
\par
If the spontaneous chiral symmetry breakdown in the infrared regime of QCD is 
governed by chiral $SU(3)_L \times SU(3)_R$ 
transformations then the effective interaction between the octet of Goldstone 
boson fields $\Phi_i$ and quarks is a pseudoscalar flavor scalar   
\begin{eqnarray}
{\cal L}_{int}=-{g_A\over 2f_\pi }\sum_{i=1}^{8}\bar q \partial_\mu 
               \gamma^\mu \gamma_5 \lambda_i \Phi_i q         
\label{lint}
\end{eqnarray}
which flips the polarization of quarks: 
$q_{\downarrow}\rightarrow q_{\uparrow}+GB$, etc. 
In Eq.~\ref{lint}, $\lambda_i$, $(i=1,2,...,8)$ are the Gell-Mann SU(3) flavor 
matrices, and $g_A$ is the dimensionless axial vector quark coupling constant 
that is taken to be 1 here, while    
\begin{equation} 
g_A^{(3)}=\Delta u -\Delta d=\Delta_3={\cal F}+{\cal D}
         =(G_A/G_V)_{n\rightarrow p}, 
\label{gan}
\end{equation}
is the isotriplet axial vector coupling constant of the weak decay of the 
neutron, and $\Delta u$, $\Delta d$ and $\Delta s$ stand for the fraction of 
proton spin carried by the u, d and s quarks, respectively. They are defined 
by the matrix elements of the singlet, triplet, octet axial vector currents, 
$A^{(i)}_{\mu}$ for i$=$0,3,8 of the nucleon state at zero momentum 
transfer. Similar axial vector matrix elements for the hyperons 
define their axial charges. It is also common to define the hypercharge spin 
fraction $\Delta_8$ and the total proton spin 
$2S_z=\Delta \Sigma$ in the infinite momentum frame as  
\begin{equation}
\Delta_8=\Delta u +\Delta d -2\Delta s=3{\cal F}-{\cal D}, \qquad 
\Delta \Sigma=\Delta u +\Delta d +\Delta s.  
\label{deli}
\end{equation}
\par
{}The success of hadronic mass relations suggests that a chiral interaction 
which breaks the SU(3) flavor symmetry also be governed by the flavor 
generator $\lambda_8$, as it is expected to originate from the mass difference 
between the strange and up and down quarks (and the corresponding mass 
differences of the Goldstone bosons).  
\par
{}Writing only the flavor dependence of these interactions we therefore extend 
the SU(3) symmetric Eq.~\ref{lint} to the standard form in~\cite{WSK},  
\begin{eqnarray}   
L_{int}= {g_8\over \sqrt 2} \sum_{i=1}^{8} \bar q (1+\epsilon \lambda_8) 
      \lambda_i \Phi_i q ,
\label{fint}
\end{eqnarray}
\begin{eqnarray}
 {1\over \sqrt 2}\sum_{i=1}^{8}\lambda_i \Phi_i =  \left( \begin{array}{c}  
{ 1\over \sqrt 2} \pi^0 + { 1\over \sqrt 6} \eta \qquad \pi^+ \qquad  K^+ \cr
  \pi^- \quad -{ 1\over \sqrt 2}\pi^0+{ 1\over \sqrt 6}\eta \quad K^0 \cr
   K^- \qquad \bar K^0 \qquad  -{2\over \sqrt 6}\eta  
\end{array}\right) . 
\label{flama}  
\end{eqnarray}
Here $g_8^2:=a \sim f_{\pi NN}^2/4\pi \approx 0.08$ where 
$f_{\pi NN}:=g_{\pi NN} m_\pi/2m_N$ denotes the pseudovector $\pi$N coupling 
constant and $g_{\pi NN}$ the pseudoscalar one. The latter can be related to 
Eq.~\ref{lint} via the pion's Goldberger-Treiman relation $g_{\pi NN}f_\pi=
g_A^{(3)}m_N$. Despite the nonperturbative nature of the chiral 
symmetry breakdown the interaction between quarks and Goldstone bosons is 
small enough for a {\bf perturbative expansion in $g_8$} to apply. Note also 
that $\epsilon$ is the SU(3) breaking parameter which was found to be small, 
$\simeq$ 0.2~\cite{WSK}, in line with the small constituent quark mass ratio 
$m_q/m_s\approx$ 0.5 to 0.6.    

\par
{}From Eq.~\ref{fint} the following transition probabilities 
$P(u_{\uparrow} \rightarrow \pi^+ + d_{\downarrow})$,...
for chiral fluctuations of quarks can be organized as coefficients in the 
symbolic reactions: 
\begin{eqnarray} u_{\uparrow} \rightarrow a(1+{\epsilon \over \sqrt{3}})^2 
   (\pi^+ + d_{\downarrow})
   +a(1+{\epsilon \over \sqrt{3}})^2 {1\over 6}(\eta +u_{\downarrow})
   +a(1+{\epsilon \over \sqrt{3}})^2 {1\over 2}(\pi^0 + u_{\downarrow})\cr
   +a(1-{2\epsilon \over \sqrt{3}})^2 (K^+ + s_{\downarrow}),\cr
 d_{\uparrow} \rightarrow a(-1-{\epsilon \over \sqrt{3}})^2 (\pi^- 
   + u_{\downarrow})
   +a(1+{\epsilon \over \sqrt{3}})^2 {1\over 6}(\eta +d_{\downarrow})
   +a(-1-{\epsilon \over \sqrt{3}})^2 {1\over 2}(\pi^0 + d_{\downarrow})\cr
   +a(1-{2\epsilon \over \sqrt{3}})^2 (K^0 + s_{\downarrow}),\cr
 s_{\uparrow} \rightarrow a(1-{2\epsilon \over \sqrt{3}})^2 {2\over 3}(\eta 
   +s_{\downarrow})+a(-1+{2\epsilon \over \sqrt{3}})^2 (K^- + 
    u_{\downarrow}) +a(-1+{2\epsilon \over \sqrt{3}})^2 (\bar K^0 +
    d_{\downarrow}),\cr  
\label{fluc}
\end{eqnarray}
and similar ones for the other quark polarization.  

\par
{}From the u and d quark lines in Eq.~\ref{fluc} the total meson emission 
probability P of the proton is given to first 
order in the Goldstone fluctuations by 
\begin{eqnarray}
P = a[{5\over 3}(1+{\epsilon \over \sqrt{3}})^2  
    +(1-{2\epsilon \over \sqrt{3}})^2],  
\label{prob}
\end{eqnarray}  
while the total strange quark probability 
\begin{eqnarray}
P_s = {8\over 3}a(1-{2\epsilon \over \sqrt{3}})^2
\label{sprob}
\end{eqnarray}
can be read off the s quark line in Eq.~\ref{fluc}.
\par
One of the main tools for incorporating chiral fluctuations in the NQM's 
valence quark spin fractions $\Delta u_v=4/3,\quad  \Delta d_v=-1/3, 
\quad \Delta s_v=0$ 
is the proton's probability composition expression~\cite{EHQ,WSK,SMW}      
\begin{eqnarray}
(1-P)({5\over 3} \hat u_{\uparrow} + {1\over 3} \hat u_{\downarrow} 
 +{1\over 3} \hat d_{\uparrow} + {2\over 3} \hat d_{\downarrow})
 + {5\over 3} P(u_{\uparrow}) + {1\over 3} P(u_{\downarrow}) 
 + {1\over 3} P(d_{\uparrow}) + {2\over 3} P(d_{\downarrow}).
\label{qp}
\end{eqnarray} 
It follows from its SU(6) spin-flavor wave function. For those other baryon 
NQM spin-flavor wave functions that can be obtained from the nucleon by  
permutations of quarks, such as
\begin{eqnarray} 
|n\rangle=-|p(u\leftrightarrow d)\rangle,\quad
|\Sigma ^+\rangle=|p(d\rightarrow s)\rangle,\quad    
|\Sigma ^-\rangle=-|n(u\rightarrow s)\rangle,\cr
|\Xi ^-\rangle=-|p(u\rightarrow s)\rangle,\quad
|\Xi ^0\rangle=|n(d\rightarrow s)\rangle,\quad 
\label{wfp}        
\end{eqnarray}    
the corresponding composition law becomes 
\begin{eqnarray}
(1-P)({5\over 3} \hat u_{\uparrow} + {1\over 3} \hat u_{\downarrow}) 
 +(1-P_s)({1\over 3} \hat s_{\uparrow} + {2\over 3} \hat s_{\downarrow})
 + {5\over 3} P(u_{\uparrow}) + {1\over 3} P(u_{\downarrow})
 + {1\over 3} P(s_{\uparrow}) + {2\over 3} P(s_{\downarrow})
\label{qph1}
\end{eqnarray} 
for the $\Sigma ^+$, 
\begin{eqnarray}
(1-P)({1\over 3} \hat d_{\uparrow} + {2\over 3} \hat d_{\downarrow}) 
 +(1-P_s)({5\over 3} \hat s_{\uparrow} + {1\over 3} \hat s_{\downarrow})
 + {1\over 3} P(d_{\uparrow}) + {2\over 3} P(d_{\downarrow})
 + {5\over 3} P(s_{\uparrow}) + {1\over 3} P(s_{\downarrow})
\label{qph2}
\end{eqnarray}
for the $\Xi ^-$, etc. Moreover, since the {\bf antiquarks from Goldstone 
bosons are unpolarized}, which is a major consequence of chiral dynamics 
and supported by the data, we use $\bar u_\uparrow = \bar u_\downarrow$ in the 
spin fractions 
$\Delta u = u_\uparrow -u_\downarrow +\bar u_\uparrow -\bar u_\downarrow$, etc. 
and $\Delta s=\Delta s_{sea}$, $\Delta \bar u=\Delta \bar d=\Delta \bar s =0$.  
Small antiquark polarizations are consistent with the most recent SMC 
data~\cite{SMC}. The NQM valence quark spin fractions in conjunction with the 
probabilities displayed in Eq.~\ref{fluc} and  Eq.~\ref{qp} to first order 
in the chiral fluctuations then yield the following spin fractions for the 
proton~\cite{WSK} 
\begin{eqnarray}
\Delta u_P=u_{\uparrow}-u_{\downarrow}={4\over 3}(1-P)
 -{5\over 9}a(1+{\epsilon\over \sqrt{3}})^2,
\label{del1}
\end{eqnarray}
\begin{eqnarray}
\Delta d_P=-{1\over 3}(1-P)-{10\over 9}a(1+{\epsilon \over \sqrt{3}})^2, 
\qquad 
\label{del2}
\end{eqnarray}
\begin{equation}
\Delta s_P=-a(1-{2\epsilon\over \sqrt{3}})^2.  \qquad
\label{del3}
\end{equation}
and the relevant hyperons
\begin{eqnarray}
\Delta u_{\Sigma ^+}={4\over 3}(1-P)
 -{8\over 9}a(1+{\epsilon\over \sqrt{3}})^2
 +{1\over 3}a(1-{2\epsilon\over \sqrt{3}})^2,
\label{del4}
\end{eqnarray}
\begin{eqnarray}
\Delta d_{\Sigma ^+}=-{4\over 3}a(1+{\epsilon \over \sqrt{3}})^2
 +{1\over 3}a(1+{\epsilon\over \sqrt{3}})^2,
\qquad 
\label{del5}
\end{eqnarray}
\begin{equation}
\Delta s_{\Sigma ^+}=-{1\over 3}(1-P_s)
 -{10\over 9}a(1-{2\epsilon\over \sqrt{3}})^2,\qquad
\label{del6}
\end{equation}

\begin{eqnarray}
\Delta u_{\Xi ^-}=-{4\over 3}a(1+{\epsilon\over \sqrt{3}})^2
 +{1\over 3}a(1-{2\epsilon\over \sqrt{3}})^2,
\label{del7}
\end{eqnarray}
\begin{eqnarray}
\Delta d_{\Xi ^-}=-{1\over 3}(1-P)
+{2\over 9}a(1+{\epsilon \over \sqrt{3}})^2
 -{4\over 3}a(1-{2\epsilon\over \sqrt{3}})^2,
\qquad 
\label{del8}
\end{eqnarray}
\begin{equation}
\Delta s_{\Xi ^-}={4\over 3}(1-P_s)
 -{5\over 9}a(1-{2\epsilon\over \sqrt{3}})^2.\qquad
\label{del9}
\end{equation}

The identities $\Delta u_{\Sigma ^-}=\Delta d_{\Sigma ^+}$, 
$\Delta d_{\Sigma ^-}=\Delta u_{\Sigma ^+}$ and 
$\Delta u_{\Xi ^0}=\Delta d_{\Xi ^-}$, 
$\Delta d_{\Xi ^0}=\Delta u_{\Xi ^-}$ follow from quark  
permutations $u\leftrightarrow d$, while the $\Delta s_B$ stay the same. 
Although these hyperon spin fractions are not measured they will serve us as a 
diagnostic tool in a study of the Karl-Sehgal formulas for magnetic moments in 
Sect. IV. We now turn our attention to another major ingredient that is still 
missing in the spin fractions, viz. relativistic effects.  

\section{ Spin-Flavor Fractions with Relativistic Effects}

Since quarks were first detected as pointlike particles at SLAC late in the 
1960s, spin fractions are measured in deep inelastic scattering (DIS) from 
the proton at high energy and momentum. In the Bjorken limit, where 
$0\leq x=-q^2/2pq\leq 1$, the relevant hadronic tensor is dominated by 
contributions from the tangent plane to the light cone. Thus, the appropriate 
form of relativistic dynamics is Dirac's light cone or front form rather than 
the more usual instant form~\cite{PAMD}. The front form is obtained from the 
instant form in the infinite momentum limit, and this amounts to a change of 
momentum variables to $p^+=p_0+p_z,\ p^-=p_0-p_z$~\cite{SK}. The longitudinal 
quark fractions are defined as $x_i=p^+_i/P^+$ with the total proton 
momentum $P^+=\sum_{i}p^+_i$ so that $\sum_{i}x_i=1$. The 
constituent quark model has been formulated on the light cone in these 
variables by many authors~\cite{LCQM,W,WK,BS}. Such LCQM's include kinematic 
boosts at the expense of interaction dependent angular momentum operators. 
Free Melosh rotations are of central importance in the LCQM's for the 
construction of relativistic many-body spin-flavor wave functions for 
hadrons from those of the NQM. In contrast, chiral bag models treat only the 
interacting quark relativistically, thereby violating translation and 
Lorentz invariance. 
\par     
For a three-quark bound state the relative four-momentum variables are the 
space-like Jacobi momenta in which the kinematic invariants $x_i$ play the 
role of masses. For example, $q_3$ is the relative quark momentum between 
the up quarks of the proton in the uds-basis and $Q_3$ between 
the down quark and the up quark pair, so that for the + and $\perp=(x,y)$ 
components 
\begin{eqnarray}
q_3={x_1p_2-x_2p_1\over x_1+x_2},\quad Q_3=(x_1+x_2)p_3-x_3(p_1+p_2), 
\label{jac}
\end{eqnarray}
etc. In light front dynamics the total momentum motion rigorously separates 
from the internal motion. Therefore, the internal baryon wave function 
$\psi(x_i, q_3, Q_3, \lambda_i) $ does not change under kinematic Lorentz 
transformations or translations. Thus, if the wave function is known in the 
baryon rest frame it is known everywhere.   
\par
In these circumstances the proton composition law of Eq.~\ref{qp} will be 
modified as follows in a relativistic quark model based on light-front 
dynamics   
\begin{eqnarray}
(1-P)(u^0_{\uparrow} \hat u_{\uparrow} +u^0_{\downarrow} \hat u_{\downarrow} 
 +d^0_{\uparrow} \hat d_{\uparrow} +d^0_{\downarrow} \hat d_{\downarrow})
 +u^0_{\uparrow} P(u_{\uparrow}) + u^0_{\downarrow} P(u_{\downarrow}) 
 +d^0_{\uparrow} P(d_{\uparrow}) + d^0_{\downarrow} P(d_{\downarrow}), 
\label{rqp}
\end{eqnarray}
involving the polarized quark-parton probabilities $q^0_{\lambda }$ in the 
LCQM~\cite{W} that are obtained from the standard quark-parton probability 
densities 
\begin{eqnarray}
q_{\lambda_i}(x)=\sum_{\lambda_j, j\neq i}\int
{dx_1 dx_2 d^2\vec q_{3\perp}d^2\vec Q_{3\perp}\over {(16\pi^3)^2}}
\delta (x_i-x)|\psi_N(x_j, \vec q_{3\perp}, \vec Q_{3\perp} \lambda_j)|^2
\label{qx}
\end{eqnarray}  
by integrating over Bjorken x.  
Proceeding as in Sect. II, Eqs.~\ref{fluc}, \ref{rqp} yield the proton spin 
fractions
\begin{eqnarray}
\Delta u_P=\Delta u^0(1-P)-({2\over 3}\Delta u^0+\Delta d^0)a
(1+{\epsilon\over \sqrt{3}})^2,
\label{rdel1}
\end{eqnarray}
\begin{eqnarray}
\Delta d_P=\Delta d^0(1-P)-({2\over 3}\Delta d^0+\Delta u^0)a
(1+{\epsilon\over \sqrt{3}})^2,
\qquad 
\label{rdel2}
\end{eqnarray}
\begin{equation}
\Delta s_P=-(\Delta u^0+\Delta d^0)a(1-{2\epsilon\over \sqrt{3}})^2,\qquad
\label{rdel3}
\end{equation}
where the $\Delta q^0=q^0_{\uparrow}-q^0_{\downarrow}$ contain the 
probabilities of Eq.\ref{rqp}. When $\Delta u^0, \Delta d^0$ are replaced by 
the NQM spin fractions, Eqs.~\ref{rdel1} to \ref{rdel3} reduce to 
Eqs.~\ref{del1} to \ref{del3}. If one ignores the difference between 
relativistic effects for the u,d quarks and the s quark, spin fractions for 
hyperons can be obtained by the relevant quark permutations from the proton 
or neutron,  
\begin{eqnarray}
\Delta u_{\Sigma ^+}=\Delta u^0(1-P)
-{2\over 3}\Delta u^0 a(1+{\epsilon\over \sqrt{3}})^2
-\Delta d^0a(1-{2\epsilon\over \sqrt{3}})^2,
\label{rdel4}
\end{eqnarray}
\begin{eqnarray}
\Delta d_{\Sigma ^+}=-\Delta u^0 a(1+{\epsilon\over \sqrt{3}})^2
-\Delta d^0 a(1-{2\epsilon\over \sqrt{3}})^2,
\qquad 
\label{rdel5}
\end{eqnarray}
\begin{equation}
\Delta s_{\Sigma ^+}=\Delta d^0 (1-P_s)
-({2\over 3}\Delta d^0+\Delta u^0)a(1-{2\epsilon\over \sqrt{3}})^2,\qquad
\label{rdel6}
\end{equation}  

\begin{eqnarray}
\Delta u_{\Xi ^-}=-\Delta d^0 a(1+{\epsilon\over \sqrt{3}})^2
-\Delta u^0a(1-{2\epsilon\over \sqrt{3}})^2,
\label{rdel7}
\end{eqnarray}
\begin{eqnarray}
\Delta d_{\Xi ^-}=\Delta d^0(1-P)
-{2\over 3}\Delta d^0 a(1+{\epsilon\over \sqrt{3}})^2
-\Delta u^0 a(1-{2\epsilon\over \sqrt{3}})^2,
\qquad 
\label{rdel8}
\end{eqnarray}
\begin{equation}
\Delta s_{\Xi ^-}=\Delta u^0 (1-P_s)
-({2\over 3}\Delta u^0+\Delta d^0)a(1-{2\epsilon\over \sqrt{3}})^2,\qquad
\label{rdel9}
\end{equation}  
and for the other hyperons by the appropriate quark permutation. The antiquark 
fractions in~\cite{WSK} stay unchanged for the proton; for the hyperons they 
are given by   
\begin{eqnarray}
\bar u_{\Sigma ^+}={2a\over 9}[4(1+{\epsilon\over \sqrt{3}})^2
+5(1-{2\epsilon\over \sqrt{3}})^2],\quad
\bar d_{\Sigma ^+}={10a\over 9}[2(1+{\epsilon\over \sqrt{3}})^2
+(1-{2\epsilon\over \sqrt{3}})^2],\cr
\bar s_{\Sigma ^+}={2a\over 9}(1+{\epsilon\over \sqrt{3}})^2
+{22\over 9}a(1-{2\epsilon\over \sqrt{3}})^2,\quad
\bar s_{\Xi ^-}={a\over 9}(1+{\epsilon\over \sqrt{3}})^2
+{17\over 9}a(1-{2\epsilon\over \sqrt{3}})^2,\cr
\bar u_{\Xi ^-}={10\over 9}a(1+{\epsilon\over \sqrt{3}})^2
+{20\over 9}a(1-{2\epsilon\over \sqrt{3}})^2,\quad
\bar d_{\Xi ^-}={4\over 9}a(1+{\epsilon\over \sqrt{3}})^2
+{20\over 9}a(1-{2\epsilon\over \sqrt{3}})^2,\cr
\label{anti2}
\end{eqnarray}
and similar expressions for the other hyperons by applying the appropriate 
quark permutations.  
\par
Finally, let us turn to the $\eta $ meson case and its recent problems.
The $\eta $ meson arises as the octet Goldstone boson when the chiral 
$SU(3)_L\times SU(3)_R$ symmetry is spontaneously broken. Predictions from 
PCAC are not in good agreement with experiments, e. g. its octet 
Goldberger-Treiman relation is violated because it implies a fairly large 
$\eta $NN
coupling constant which disagrees with the much smaller value extracted from 
analyses of both $p\bar p$ collisions~\cite{pbar} and recent precision data 
from MAMI~\cite{MAMI} on $\eta $ photoproduction off the proton at threshold. 
Corrections from chiral perturbation theory are of order 30$\%$~\cite{SW} and 
much too small to help one understand the problem of the suppressed $\eta $NN 
coupling better. This conflict with the data can be avoided if the 
$\eta $ meson couples only to the strange, but not the up and down, quarks 
and takes on features of a strange Goldstone boson~\cite{KW}. This amounts 
to striking the $\eta $ meson from the up and down reactions in Eq.~\ref{fluc} 
so that the factor 5/3 in P changes to 3/2, 
${2\over 3}\Delta u^0+\Delta d^0$ becomes ${1\over 2}\Delta u^0+\Delta d^0$, 
and ${2\over 3}\Delta d^0+\Delta u^0$ goes to 
${1\over 2}\Delta d^0+\Delta u^0$ in the expressions for the spin fractions.   
This case is labeled $\eta ^{(s)}$ in the Tables 1 to 7, while the standard 
octet case is labeled $\eta ^{(8)}$; it is included here not to obtain a 
better fit but because it may turn out to be more realistic, as it avoids a  
conflict with the $g_{\eta NN}$ data. In this case, $\bar u/\bar d$ is lowered 
from the SU(3) symmetric value 3/4 to 7/11. Note that the experimental value 
0.51$\pm$0.04(stat.)$\pm$0.05(syst.) is at x$=$0.18~\cite{NA51} and not summed 
over Bjorken x.    

\section{ Magnetic Moments of Baryons}
\par
{}In the nonrelativistic quark model with three-valence quark wave functions 
for the baryon octet the magnetic moments contain the NQM spin fractions 
$\Delta u=$4/3, $\Delta d=$-1/3, $\Delta s=$0, e.g. 
$\mu (p)={4\over 3}\mu _u-{1\over 3}\mu _d$, etc. 
For other octet baryon flavor wave functions that follow from the proton or 
neutron by the quark permutations of Eq.~\ref{wfp}, the simple NQM expressions 
for the nucleon's magnetic moments suggest the validity of the Karl-Sehgal 
equations~\cite{KS}, called K-S below,  
\begin{eqnarray}
\mu (p)=\mu_ u\Delta u+\mu _d\Delta d+\mu _s\Delta s, \qquad
\mu (n)=\mu_ u\Delta d+\mu _d\Delta u+\mu _s\Delta s,\cr
\mu (\Sigma ^+)=\mu_ u\Delta u+\mu _d\Delta s+\mu _s\Delta d, \qquad
\mu (\Sigma ^-)=\mu_ u\Delta s+\mu _d\Delta u+\mu _s\Delta d,\cr
\mu (\Xi ^-)=\mu_ u\Delta s+\mu _d\Delta d+\mu _s\Delta u, \qquad
\mu (\Xi ^0)=\mu_ u\Delta d+\mu _d\Delta s+\mu _s\Delta u, 
\label{KSR}
\end{eqnarray}
which are SU(3) symmetric when effective quark magnetic moments 
are chosen in accord with the quark charge ratios, $-2\mu _d=\mu _u$,
$\mu _d= \mu _s$.  
Let us study their validity when chiral fluctuations are included 
by comparing with the appropriate additive baryon magnetic moments
\begin{equation}
\mu (B)=\mu _u\Delta u_B+\mu _d\Delta d_B+\mu _s\Delta s_B
\label{bmm}
\end{equation} 
with $\Delta q_B$ from Eqs.~\ref{del4},..,\ref{del9}. 
To this end we write linear expressions for $\Delta q_B= 
c_1\Delta u_P+c_2\Delta d_P+c_3\Delta s_P$ in terms of the $\Delta q_P$ of the 
proton in Eqs.~\ref{del1},\ref{del2},\ref{del3} and obtain 
\begin{eqnarray}
\Delta u_{\Sigma ^+}={16\over 15}\Delta u_P+{4\over 15}\Delta d_P
-{1\over 3}\Delta s_P,\quad
\Delta d_{\Sigma ^+}={4\over 15}\Delta u_P+{16\over 15}\Delta d_P
-{1\over 3}\Delta s_P,\cr
\Delta s_{\Sigma ^+}=-{1\over 9}\Delta u_P+{5\over 9}\Delta d_P
 +{5\over 9}\Delta s_P,\quad
\Delta s_{\Sigma ^-}=-{1\over 9}\Delta u_P+{5\over 9}\Delta d_P
+{5\over 9}\Delta s_P,\cr
\Delta u_{\Sigma ^-}={4\over 15}\Delta u_P+{16\over 15}\Delta d_P
-{1\over 3}\Delta s_P,\quad
\Delta d_{\Sigma ^-}={16\over 15}\Delta u_P+{4\over 15}\Delta d_P
-{1\over 3}\Delta s_P,\cr
\Delta u_{\Xi ^-}=-{1\over 15}\Delta u_P-{4\over 15}\Delta d_P
+{4\over 3}\Delta s_P,\quad
\Delta d_{\Xi ^-}=-{4\over 15}\Delta u_P-{1\over 15}\Delta d_P
+{4\over 3}\Delta s_P,\cr
\Delta s_{\Xi ^-}={4\over 9}\Delta u_P-{20\over 9}\Delta d_P
+{25\over 9}\Delta s_P,\quad
\Delta s_{\Xi ^0}={4\over 9}\Delta u_P-{20\over 9}\Delta d_P
+{25\over 9}\Delta s_P,\cr
\Delta u_{\Xi ^0}=-{4\over 15}\Delta u_P-{1\over 15}\Delta d_P
+{4\over 3}\Delta s_P,\quad
\Delta d_{\Xi ^0}=-{1\over 15}\Delta u_P-{4\over 15}\Delta d_P
+{4\over 3}\Delta s_P.\cr
\label{lsr}
\end{eqnarray}
Clearly, all coefficients are independent of $a$ and $\epsilon$, and 
the K-S relations are not satisfied because chiral fluctuations 
break the SU(3) flavor symmetry so that, e. g., the total chiral 
probabilities $P\neq P_s$, etc. However, setting 
\begin{eqnarray}
\Delta u_{\Sigma ^+}=\Delta u_P,\quad 
\Delta d_{\Sigma ^+}=\Delta s_P,\quad 
\Delta s_{\Sigma ^+}=\Delta d_P,\cr
\Delta u_{\Sigma ^-}=\Delta s_P,\quad 
\Delta d_{\Sigma ^-}=\Delta u_P,\quad 
\Delta s_{\Sigma ^-}=\Delta d_P,\cr
\Delta d_{\Xi ^-}=\Delta d_P,\quad
\Delta u_{\Xi ^-}=\Delta s_P,\quad  
\Delta u_{\Xi ^-}=\Delta s_P,\cr
\Delta u_{\Xi ^0}=\Delta d_P,\quad
\Delta d_{\Xi ^0}=\Delta s_P,\quad
\Delta s_{\Xi ^0}=\Delta u_P,
\label{ksf}
\end{eqnarray}
according to the K-S relations in conjunction with Eq.~\ref{bmm}, and solving 
for $\Delta s_P$ yields just one constraint 
$\Delta s_P=[\Delta u_P+4\Delta d_P]/5$ which is small, O(a), but too 
large by a factor 
$(1+\epsilon/\sqrt{3})^2/(1-2\epsilon/\sqrt{3})^2\approx$15/4 for 
$\epsilon=$1/3 compared to the correct $\Delta s_P$.  
\par
These linear relations can be generalized to relativistic quark models. The 
coefficients of such generalized linear relations then depend only on 
$\Delta u^0, \Delta d^0$ of the LCQM. Again, they all lead to a single 
constraint for $\Delta s_P$ which differs from the correct $\Delta s_P$ by 
the same $\epsilon $ dependent factor.  
\par
Next we show that the K-S relations are no longer valid when relativistic 
effects are included which break the SU(3) flavor symmetry also, but in 
different ways. In~\cite{WK} 
the nucleon magnetic moments $\mu (p)=$2.80 n.m., $\mu (n)=$-1.73 n.m. for 
quark mass $m_u=m_d=m_q=$0.33 GeV and harmonic 
oscillator parameter $\alpha =$0.32 GeV were obtained in the LCQM using Dirac 
magnetic moments for the quarks. The valence quark spin fractions of the 
LCQM are $\Delta u=$0.96 and $\Delta d=$-0.24 for this case so that 
$g^{(3)}_A=$1.2. 
The K-S formulas yield instead 
\begin{eqnarray}
\mu (p)=(2\Delta u-\Delta d){m_N\over 3m_q}=2.05\ n.m.,\quad   
\mu (n)=(2\Delta d-\Delta u){m_N\over 3m_q}=-1.36\ n.m.
\label{mks}
\end{eqnarray}
The discrepancy is smaller for the lower quark mass $m_q=$0.263 GeV adopted 
in the LCQM~\cite{BS} where $\mu (p)=$2.81 n.m. and $\mu (n)=$-1.66 n.m. are 
calculated, while $\Delta u=$1 and $\Delta d=$-1/4 of this LCQM yield the 
K-S magnetic moments $\mu (p)=$2.68 n.m., $\mu (n)=$\quad -1.78 n.m. The small 
difference between these values and the magnetic moments~\cite{WK} above 
illustrates that in quark models which include relativistic effects to all 
orders in $p/m_q$ or v/c, magnetic moments of baryons are much less 
sensitive to the quark mass than the $1/m_q$ dependence of the K-S formulas 
would suggest, so that additive formulas like Eq.~\ref{bmm} become invalid 
also. Therefore, it is easy to reach misleading conclusions from 
fits of K-S formulas~\cite{S} to the data. Finally, let us mention that 
electromagnetic gauge invariance requires many-body quark currents to be 
present~\cite{Lev} which invalidate additive quark model results for magnetic 
moments also. Pion pair and exchange currents have been studied~\cite{WW} and 
found to be not negligible, but there are significant cancellations of such 
pion loop contributions to the nucleon magnetic moments so that their net 
effect is just a 1 to 2 $\%$ correction.   

\section{Numerical Results}
\par
The simplest relativistic quark models depend on two parameters, the common 
u, d quark mass $m_q$ and the harmonic oscillator constant $\alpha $ of the 
confinement potential. The proton size determines $\alpha $ so that  
$1/\alpha \sim \langle r^2\rangle_P^{1/2}$ up to relativistic corrections. 
Since the proton magnetic moment is not changed much by chiral 
fluctuations, in contrast to the axial charge $g^{(3)}_A$, we determine the 
range of the parameter $\alpha $ from a wave function independent relation for 
$\mu _p$ as a function of proton radius~\cite{BS}. 
\par  From Fig.1 in~\cite{BS} we find 
$\alpha \approx 3.6/m_N\sim$0.26 GeV $\sim(0.76 fm)^{-1}$. For $\alpha =$0.25 
GeV, in Table 1 proton spin and flavor fractions for the NQM are compared with 
those of the chiral NQM and chiral LCQM of~\cite{WK}. The spin fractions of 
the LCQM are $\Delta u^0=$1.1, $\Delta d^0=$-0.275, $\Delta s^0=$0, so that 
$g^{(3)}_A=$1.375, for $m_q=$0.33 GeV. Typically, $\Delta s$ values are very 
small, and $\Delta u$, $\Delta \Sigma $ and $g^{(3)}_A$ are substantially 
reduced from the NQM and LCQM values by chiral fluctuations.  
A comparison of the relativistic and nonrelativistic results in Table 2 shows 
that relativistic effects lead to a lower chiral strength $a$, but the SU(3) 
flavor breaking parametrized in terms of $\epsilon $ stays unchanged in the 
octet $\eta $ cases, while $\epsilon $ is reduced for the nonrelativistic 
$\eta ^{(s)}$ case and enhanced for the relativistic $\eta ^{(s)}$ case. Thus, 
the effective SU(3) breaking brought about by relativistic effects is  
sensitive to the role of the $\eta $ meson in the chiral dynamics.  

\par These spin and flavor fractions are the chiral quark model 
results of Sects. III and IV that are independent of momentum Q$^2$ and valid 
at long distances below the chiral scale $\Lambda _{\chi }=4\pi f_{\pi}$.  
In the chiral limit, the axial charges of the nucleon (and the baryon octet), 
i.e. $\Delta _3=g^{(3)}_A$ and $\Delta _8$, are constants independent of 
Q$^2$ because conserved currents have vanishing anomalous dimension. The 
singlet axial charge, $\Delta \Sigma $, is not conserved because of the 
axial anomaly, but its Q$^2$ dependence is rather weak to lowest orders in 
pQCD. Due to the U(1) anomaly of the singlet axial vector current the axial 
charge contains a gluon contribution. In the Adler-Bardeen renormalization 
scheme~\cite{AB} the Q$^2$ dependent axial charges are given by 
$a_i(Q^2)=\Delta q_i-{\alpha _s(Q^2)\over 2\pi }\Delta g(Q^2)$ for i$=$u,d,s 
and $a_0(Q^2)=\Delta \Sigma-n_f{\alpha _s(Q^2)\over 2\pi }\Delta g(Q^2)$ 
for $n_f=$3 flavors. At Q$^2=$5 GeV$^2$, the SMC~\cite{SMC} experiment has  
determined $\Delta g=$1.7$\pm$1.1  so that 
${\alpha _s(Q^2)\over 2\pi }\Delta g(Q^2)\approx 0.084\pm0.054$. These Q$^2$ 
dependent axial charges $a_i$ are compared with the data in Table 2. 
Clearly, the nonrelativistic and relativistic quark model results are in fair 
agreement with the data, especially for the $\eta ^{(s)}$ case. Typically 
$g^{(3)}_A$ for the NQM cases is above the data and for the LCQM below the 
data, but not by much. Without the momentum dependent gluon contribution the   
relativistic fits become worse, and nonrelativistic fits better, comparable to 
those in Table 4. In this sense the gluon contribution supports the 
relativistic chiral quark models that are 
expected to contain more of the relevant physics than the chiral NQM. However, 
the results in Table 2 are not the best fits. In Table 3 results are presented 
for the LCQM for a smaller value of $\alpha $ that produce the better fits in 
Table 4. This parameter range is 
driven by the proton's axial charge $g^{(3)}_A$. Fitting the axial charge with 
relativistic effects and chiral fluctuations is more difficult than in the 
nonrelativistic case, especially without the singlet axial gluon contribution,  
requiring a low value of $\alpha $ so as to minimize the reduction of 
$g^{(3)}_A$ by relativistic effects.\footnote{Quark models that do not include 
chiral dynamics typically lead to lower $m_q$ and smaller proton size, but are 
misleading because they are inconsistent with chiral perturbation theory.} 
Thus, even though $g^{(3)}_A$ has the large value 1.375 for the 
relativistic valence quark model in Tables 1, 2 despite relativistic effects, 
the axial charge still falls below the observed value when chiral fluctuations 
are included. The preference of LCQMs for an unrealistically large proton size 
$\geq$ 1 fm should not be taken too literally. Given the simplicity of the 
models one should not expect better fits than those in Table 2, and it is not 
surprising, therefore, that much better fits like those in Table 4 push the 
parameter $\alpha $ to extreme values.    
\par
We have also examined a light-cone quark model~\cite{W} that includes two 
harmonic oscillator constants $\alpha ^2(1\pm D)$ in the Gaussian momentum 
wave function which, for D$>$0, amounts to smaller u-d quark pairs in the 
proton simulating an attractive spin force between the u, d quark pairs. The 
value D$=$0.37 generates the full $\Delta $(1232)-N mass splitting. The spin 
and flavor fractions of the proton do not change by much, nor do the SU(3) 
flavor breaking parameters $\epsilon $ and $a$, when this quark clustering is 
included, nor does it help with the axial charge. Therefore, no detailed 
results are presented.  

\par
In Table 5 the magnetic moments are obtained from the additive 
relation, Eq.~\ref{bmm}, by fitting the magnetic 
moment $\mu _u$ of the up quark while maintaining fixed ratios 
$\mu _d=-{1\over 2}\mu _u$ suggested by the quark charge ratio and 
$\mu _s={3\over 5}\mu _d$ by the quark mass ratio.  
As is shown in Table 3, the magnetic moments from Eq.~\ref{bmm} give  
better fits, mainly due to the $\Xi $, than the K-S relations which the 
additive formula of Eq.~\ref{bmm} improves by properly taking SU(3) breaking 
into account. As expected, the relativistic version of the one-body relation 
in Eq.~\ref{bmm} does not improve the fits of magnetic moments of the NQM 
version noticeably, so that only one case, the LCQM results, are presented. 
The fits are better than expected on theoretical grounds discussed in Sect.IV.  
\par 
The spin fractions for the hyperons in Tables 6 and 7 show changes similar 
to the proton with chiral fluctuations. In a comparison with the spin 
fractions at the one-loop level of chiral perturbation theory~\cite{SW} we 
note that relativistic effects typically improve the dominant $\Delta q$, 
but not all three. 

\section{Summary and Conclusion}
When relativistic effects are included along with chiral dynamics to lowest 
order, it becomes more difficult for chiral quark models to reproduce the 
measured value of the axial charge $g^{(3)}_A$ of the nucleon. The range of 
relativistic quark model parameters giving acceptable fits is pushed towards 
higher u,d quark mass and larger proton size when chiral fluctuations are 
included. Chiral fluctuations that imply unpolarized antiquark fractions give 
a remarkably successful description of the spin fractions of the proton. 
The SU(3) breaking is significant. The NQM is able to simulate the missing 
relativistic effects and the axial anomaly by adjusting the chiral strength 
$a$ and the SU(3) breaking $\epsilon $, the latter being too small and 
misleading.         
\par
The extension of spin fractions to hyperons allows us to calculate their  
magnetic moments and improve the Karl-Sehgal formulas. We find that SU(3) 
breaking via chiral dynamics and relativistic effects invalidates the K-S 
formulas; and this breakdown casts doubt on them as a useful tool for 
probing the spin fractions of the proton via hyperon magnetic moments.   
\par   
Despite being more sophisticated than the NQM, relativistic chiral quark 
models are not yet able to predict the Bjorken-x dependence of the flavor and 
spin distributions of baryons. Nonetheless, the 
successful description of most spin and flavor fractions with just a few 
parameters is encouraging and suggests further development and study of these 
models.   
\section{Acknowledgement}
It is a pleasure to thank Xiaotong Song for his interest and careful reading 
of the manuscript. 

\vfill\eject

{\bf Table 1} Quark Spin and Sea Observables of the Proton; $m_q=$0.33 
GeV, $\alpha =$0.25 GeV 

$$
\offinterlineskip \tabskip=0pt 
\vbox{ 
\halign to 1.0\hsize 
   {\strut
  \vrule#                          
   \tabskip=0pt plus 30pt
 & \hfil #  \hfil                  
 & \vrule#                         
 & \hfil #  \hfil                  
 & \vrule#                         
 & \hfil #  \hfil                  
 & \vrule#                         
 & \hfil #  \hfil                  
 & \vrule#                         
 & \hfil #  \hfil                  
 & \vrule#                         
 & \hfil #  \hfil                  
   \tabskip=0pt                    %
 & \vrule#                         
  \cr                             
\noalign{\hrule}
&  && NQM && $a=$0.09 && $a=$0.1 && $a=$0.07 && $a=$0.07 &\cr
& && && $\eta ^{(8)}$ && $\eta ^{(s)}$ && $\eta ^{(8)}$ && $\eta ^{(s)}$ &\cr
& && && NQM && NQM && LCQM && LCQM &\cr
& &&  && $\epsilon =$0.32 && $\epsilon =$0.3 && 
$\epsilon =$0.32 && $\epsilon =$0.34 &\cr
\noalign{\hrule}
& $\Delta u$ && 4/3 && 0.94 && 0.96 && 0.84 && 0.88 &\cr
& $\Delta d$ && -1/3 && -0.39 && -0.41 &&-0.31 &&-0.32 &\cr
& $\Delta s$ && 0 && -0.04 && -0.04 &&-0.02 &&-0.02 &\cr
& $\Delta \Sigma $ && 1 && 0.51 && 0.50 && 0.51 && 0.53 &\cr
& $\Delta_3/\Delta_8$ && 5/3 && 2.16 && 2.17 && 2.00 && 2.01 &\cr
& $g^{(3)}_A$  && 5/3 && 1.33 && 1.37 &&1.16 && 1.2 &\cr
& ${\cal F}/{\cal D}$ && 2/3 && 0.58 && 0.58 &&0.60 &&0.60 &\cr
& $\bar u/\bar d$ && - && 3/4 && 7/11 && 3/4 && 7/11 &\cr
& $f_3/f_8$  && 1/3 && 0.22 && 0.19 && 0.24 &&0.22 &\cr
& $I_G$      && 1/3 && 0.28 && 0.24 && 0.29 &&0.27 &\cr
\noalign{\hrule}
}}$$
\vfill\eject

{\bf Table 2} Quark Spin Observables of the Proton; $m_q=$0.33 
GeV, $\alpha =$0.25 GeV, with 
\qquad singlet axial gluon contribution  

$$
\offinterlineskip \tabskip=0pt 
\vbox{ 
\halign to 1.0\hsize 
   {\strut
  \vrule#                          
   \tabskip=0pt plus 30pt
 & \hfil #  \hfil                  
 & \vrule#                         
 & \hfil #  \hfil                  
 & \vrule#                         
 & \hfil #  \hfil                  
 & \vrule#                         
 & \hfil #  \hfil                  
 & \vrule#                         
 & \hfil #  \hfil                  
 & \vrule#                         
 & \hfil #  \hfil                  
 & \vrule#                         
 & \hfil #  \hfil                  
 & \vrule#                         
 & \hfil #  \hfil                  
 & \vrule#                         
 & \hfil #  \hfil                  
 & \vrule#                         
 & \hfil #  \hfil                  
   \tabskip=0pt                    %
 & \vrule#                         
  \cr                             
\noalign{\hrule}
& \quad \quad && \quad Data \quad &&\quad $a=$0.09 \quad &&\ $a=$0.10  
&& \qquad $a=$0.07 \qquad && \qquad $a=$0.07 \qquad &\cr
& && E143\cite{E143} && $\eta ^{(8)}$ && $\eta ^{(s)}$ && $\eta ^{(8)}$ && 
$\eta ^{(s)}$ &\cr
& && at 3 GeV$^2$ && NQM && NQM && LCQM && LCQM &\cr
& && SMC\cite{SMC} && $\epsilon =$0.32 && $\epsilon =$0.3 && 
$\epsilon =$0.32 && $\epsilon =$0.34 &\cr
& && at 5 GeV$^2$ && && && && &\cr
\noalign{\hrule}
& $a_u$ && 0.84$\pm$0.05 && 0.85 && 0.87 && 0.76 && 0.80 &\cr
&       && 0.82$\pm$0.02 &&  &&  && && &\cr
& $a_d$ && -0.43$\pm$0.05 && -0.48 && -0.50 && -0.40 && -0.41 &\cr
&       && -0.43$\pm$0.02 && && && && &\cr
& $a_s$ && -0.08$\pm$0.05 && -0.12 && -0.13 && -0.11 && -0.11 &\cr
&        && -0.10$\pm$0.02 &&  &&  && && &\cr    
& $a_0$ && 0.30$\pm$0.06 && 0.26 && 0.25 && 0.26 && 0.28 &\cr
&       && 0.29$\pm$0.06 && && && && &\cr       
& \quad $a_3/a_8$ && 2.09$\pm$0.13 && 2.16 && 2.17 && 2.00 && 2.01 &\cr
& $g^{(3)}_A$  &&\quad 1.2573$\pm$0.0028 && 1.33 && 1.37 && 1.16 && 1.20 &\cr
& \quad ${\cal F}/{\cal D}$ && 0.575$\pm$0.016 && 0.58 && 0.58 && 0.60 && 0.60 
&\cr
& $\bar u/\bar d$ && 0.51$\pm$0.09 && 3/4 && 7/11 && 3/4 && 7/11 &\cr
& \quad $f_3/f_8$  && 0.23$\pm$0.05 && 0.22 && 0.19 && 0.24 && 0.22 &\cr
& $I_G$      && 0.235$\pm$0.026 && 0.28 && 0.24 && 0.29 && 0.27 &\cr
\noalign{\hrule}
}}$$
 
\vfill\eject

{\bf Table 3} Quark Spin and Sea Observables of the Proton; $m_q=$0.33 GeV,
$\alpha =$0.20 GeV

$$
\offinterlineskip \tabskip=0pt 
\vbox{ 
\halign to 1.0\hsize 
   {\strut
  \vrule#                          
   \tabskip=0pt plus 30pt
 & \hfil #  \hfil                  
 & \vrule#                         
 & \hfil #  \hfil                  
 & \vrule#                         
 & \hfil #  \hfil                  
 & \vrule#                         
 & \hfil #  \hfil                  
 & \vrule#                         
 & \hfil #  \hfil                  
 & \vrule#                         
 & \hfil #  \hfil                  
   \tabskip=0pt                    %
 & \vrule#                         
  \cr                             
\noalign{\hrule}
&  && \quad NQM \quad && \quad LCQM \quad && \qquad $a=$0.07 \qquad && 
\qquad $a=$0.08 \qquad &\cr
& && && \quad $m_q=$0.33 GeV \quad && $\eta ^{(8)}$ && $\eta ^{(s)}$ &\cr
& && && \quad $\alpha =$0.20 GeV \quad && $\epsilon =$0.38 && $\epsilon =$0.38 
&\cr
\noalign{\hrule}
& $\Delta u$ && 4/3 && 1.18 && 0.90 && 0.90 &\cr
& $\Delta d$ && -1/3 && -0.29 && -0.34 && -0.36 &\cr
& $\Delta s$ && 0 && 0 && -0.02 && -0.02 &\cr
& $\Delta \Sigma $ && 1 && 0.88 && 0.54 && 0.52 &\cr
& \qquad $\Delta_3/\Delta_8$ \qquad && 5/3 && 5/3 && 2.07 && 2.13 &\cr
& $ g^{(3)}_A$ && 5/3 && 1.47 && 1.24 && 1.26 &\cr
& ${\cal F}/{\cal D}$ && 2/3 && 2/3 && 0.59 && 0.58 &\cr
&\quad $\bar u/\bar d$ \quad &&  &&  && 3/4 && 7/11 &\cr
& \quad $f_3/f_8$ \quad && 1/3  &&  && 0.23 && 0.20 &\cr
& $I_G$      && 1/3 &&  && 0.29 && 0.25 &\cr
\noalign{\hrule}
}}$$

\vfill\eject

{\bf Table 4} Quark Spin and Sea Observables of the Proton with 
singlet axial gluon contribution, $m_q=$0.33 GeV, $\alpha =$0.2 GeV 

$$
\offinterlineskip \tabskip=0pt 
\vbox{ 
\halign to 1.0\hsize 
   {\strut
  \vrule#                          
   \tabskip=0pt plus 30pt
 & \hfil #  \hfil                  
 & \vrule#                         
 & \hfil #  \hfil                  
 & \vrule#                         
 & \hfil #  \hfil                  
 & \vrule#                         
 & \hfil #  \hfil                  
 & \vrule#                         
 & \hfil #  \hfil                  
 & \vrule#                         
 & \hfil #  \hfil                  
 & \vrule#                         
 & \hfil #  \hfil                  
   \tabskip=0pt                    %
 & \vrule#                         
  \cr                             
\noalign{\hrule}
&  && \quad Data \quad && \qquad Data \qquad && \qquad $a=$0.07 \qquad && 
\qquad $a=$0.08 \qquad &\cr
&  && E143\cite{E143} && SMC\cite{SMC} && $\eta ^{( 8)}$ && $\eta ^{(s)}$ &\cr
& &&at 3 GeV$^2$ && \quad at 5 GeV$^2$ \quad &&$\epsilon =$0.38 && 
$\epsilon =$0.38 &\cr
\noalign{\hrule}
& $a_u$ && 0.84$\pm$0.05 && 0.82$\pm$0.06 && 0.81 && 0.82 &\cr
& $a_d$ && -0.43$\pm$0.05 && -0.44$\pm$0.06 &&-0.42 && -0.44 &\cr
& $a_s$ && -0.08$\pm$0.05 && -0.10$\pm$0.06 && -0.10 && -0.11 &\cr
& $a_0$ && 0.30$\pm$0.06 && 0.28$\pm$0.17 && 0.29 && 0.27 &\cr
& \quad $a_3/a_8$ \quad && 2.09$\pm$0.13 && 2.17$\pm$0.16 && 2.07 && 2.13 &\cr
& $ g^{(3)}_A$ && \quad 1.2573$\pm$0.0028 \quad &&  && 1.24 && 1.26 &\cr
& ${\cal F}/{\cal D}$ && 0.575$\pm$0.016 &&  && 0.59 && 0.58 &\cr
& $\bar u/\bar d$ && 0.51$\pm$0.09 &&  && 3/4 && 7/11 &\cr
& \quad $f_3/f_8$ \quad && 0.23$\pm$0.05  &&  && 0.23 && 0.20 &\cr
& $I_G$      && 0.235$\pm$0.026 &&  && 0.29 && 0.25 &\cr
\noalign{\hrule}
}}$$

\vfill\eject

{\bf Table 5}\qquad Magnetic Moments of Baryons, $\alpha =$0.25 GeV, 
$m_q=$0.33 GeV    

$$
\offinterlineskip \tabskip=0pt 
\vbox{ 
\halign to 1.0\hsize 
   {\strut
  \vrule#                          
   \tabskip=0pt plus 30pt
 & \hfil #  \hfil                  
 & \vrule#                         
 & \hfil #  \hfil                  
 & \vrule#                         
 & \hfil #  \hfil                  
 & \vrule#                         
 & \hfil #  \hfil                  
 & \vrule#                         
 & \hfil #  \hfil                  
   \tabskip=0pt                    %
 & \vrule#                         
  \cr                             
\noalign{\hrule}
& magnetic &&\quad Data && K-S &&$a=$0.07 && $a=$0.07 &\cr
& moments  &&  &&   && $\eta ^{(8)}$ && $\eta ^{(s)}$ &\cr
& $\mu $[n.m.] && PDG\cite{PDG} && && $\epsilon =$0.32 && $\epsilon =$0.34 &\cr
\noalign{\hrule}
& p && 2.793 && 2.720 && 2.720 && 2.617 &\cr
& n && -1.913 && -1.965 && -1.965 && -1.890 &\cr
& $\Sigma ^+$ && 2.458$\pm$0.01 && 2.563 && 2.590 && 2.484 &\cr
& $\Sigma ^-$ && -1.160$\pm$0.025&& -0.949 && -0.949 && -0.909 &\cr
& $\Xi ^-$ &&-0.6507$\pm$0.0025 && -0.324 && -0.507 && -0.460 &\cr
& $\Xi ^0$  && -1.250$\pm$0.014 && -1.496 && -1.444 && -1.361 &\cr
\noalign{\hrule}
& $\mu _u$[n.m.] && input && 2.7 && 2.7 && 2.5 &\cr
\noalign{\hrule}
}}$$

\vfill\eject

{\bf Table 6}\qquad Quark Spin-Sea Observables of the $\Sigma ^+$; 
$m_q=$0.33 GeV, $\alpha =$0.25 GeV

$$
\offinterlineskip \tabskip=0pt 
\vbox{ 
\halign to 1.0\hsize 
   {\strut
  \vrule#                          
   \tabskip=0pt plus 30pt
 & \hfil #  \hfil                  
 & \vrule#                         
 & \hfil #  \hfil                  
 & \vrule#                         
 & \hfil #  \hfil                  
 & \vrule#                         
 & \hfil #  \hfil                  
 & \vrule#                         
 & \hfil #  \hfil                  
 & \vrule#                         
 & \hfil #  \hfil                  
   \tabskip=0pt                    %
 & \vrule#                         
  \cr                             
\noalign{\hrule}
& && Chiral && $a=$0.09 && $a=$0.10 && $a=$0.07 && $ a=$0.07 &\cr
& && Pert. && $\eta ^{(8)}$ && $\eta ^{(s)}$&& $\eta ^{(8)}$&&$\eta ^{(s)}$&\cr
& && Theory && NQM && NQM && LCQM && LCQM &\cr
& &&\quad \cite{SW} &&$\epsilon =$0.32 && $\epsilon =$0.30&&$\epsilon =$0.32&&
 $\epsilon =$0.34 &\cr
\noalign{\hrule}
& $\Delta u$ && 0.68$\pm$0.12 && 0.91 && 0.92 && 0.83 && 0.86 &\cr
& $\Delta d$ && 0.05$\pm$0.12 &&-0.16 && -0.17 &&-0.10 &&-0.10 &\cr
& $\Delta s$ && -0.49$\pm$0.09 &&-0.34 && -0.34 &&-0.28 &&-0.28 &\cr
& $\bar u$ && - && 0.17 && 0.12 && 0.13 && 0.08 &\cr
& $\bar d$ && - && 0.32 && 0.39 && 0.25 && 0.28 &\cr
& $\bar s$  && - && 0.03 && 0.10 && 0.02 && 0.06 &\cr
\noalign{\hrule}
}}$$

{\bf Table 7}\qquad Quark Spin-Sea Observables of the $\Xi ^-$; $m_q=$0.33 GeV,
$\alpha =$0.25 GeV

$$
\offinterlineskip \tabskip=0pt 
\vbox{ 
\halign to 1.0\hsize 
   {\strut
  \vrule#                          
   \tabskip=0pt plus 30pt
 & \hfil #  \hfil                  
 & \vrule#                         
 & \hfil #  \hfil                  
 & \vrule#                         
 & \hfil #  \hfil                  
 & \vrule#                         
 & \hfil #  \hfil                  
 & \vrule#                         
 & \hfil #  \hfil                  
 & \vrule#                         
 & \hfil #  \hfil                  
   \tabskip=0pt                    %
 & \vrule#                         
  \cr                             
\noalign{\hrule}
& && Chiral &&$a=$0.09 && $a=$0.10 &&$a=$0.07 && $a=$0.07 &\cr
& && Pert. && $\eta ^{(8)}$ && $\eta ^{(s)}$ && $\eta ^{(8)}$ && $\eta ^{(s)}$ 
&\cr
& && Theory && NQM && NQM && LCQM && LCQM &\cr
& &&\quad \cite{SW} &&$\epsilon =$0.32 &&$\epsilon =$0.30&&$\epsilon =$0.32&&
$\epsilon =$0.34&\cr
\noalign{\hrule}
& $\Delta u$ && -0.18$\pm$0.14 && -0.006 && -0.01 && -0.004 && -0.001 &\cr
& $\Delta d$ && -0.50$\pm$0.10 && -0.27 && -0.28 &&-0.24 &&-0.24 &\cr
& $\Delta s$ && 0.83$\pm$0.12 && 1.19 && 1.16 && 1.01 && 1.01 &\cr
& $\bar u$ && - && 0.22 && 0.27 && 0.17 && 0.18 &\cr
& $\bar d$ && - && 0.08 && 0.13 && 0.06 && 0.08 &\cr
& $\bar s$ && - && 0.07 && 0.08 && 0.05 && 0.05 &\cr
\noalign{\hrule}
}}$$

\vfill\eject


\begin{thebibliography}{9}
\bibitem{CPT} H.\ Leutwyler, Ann.\ Phys.(N.Y.)\ {\bf 235}, 165 (1994). 
\bibitem{WS} S.\ Weinberg, Physica {\bf 96A}, 327 (1979), Phys.\ Rev.\ 
Lett.\ {\bf 65}, 1181 (1990), ibid.\ {\bf 67}, 3473 (1991).
\bibitem{DJ} R.\ Dashen and A.\ Manohar, Phys.\ Lett.\ {\bf B315}, 425 
(1993); E.\ Jenkins, ibid.\ {\bf B315}, 431, 441, 447 (1993).    
\bibitem{EHQ} E.\ J.\ Eichten, I.\ Hinchcliffe and C.\ Quigg, Phys.\ Rev.\ 
{\bf D45}, 2269 (1992).
\bibitem{Go}  K.\ Gottfried, Phys.\ Rev.\ Lett.\ {\bf 18}, 1174 (1967).
\bibitem{E143} P.\ L.\ Anthony {\it{et al.}},\ Phys.\ Rev.\ Lett. {\bf 71}, 
959 (1993); K.\ Abe {\it{et al.}}, \ Phys.\ Rev.\ Lett. {\bf 74}, 346 (1995); 
ibid.\ {\bf 75}, 25 (1995). 
\bibitem{SMC} B.\ Adeva {\it{et al.}},\ Phys.\ Lett. {\bf B302}, 553 (1993); 
ibid. {\bf B320}, 400 (1994); ibid. {\bf 369}, 93 (1996); D.\ Adams 
{\it{et al.}}, \ Phys.\ Lett. {\bf B329}, 399 (1994); ibid. {\bf B336},125 
(1994); D.\ Adams {\it{et al.}}, hep-ex/9702005.  
\bibitem{WSK} H.\ J.\ Weber, X.\ Song, and M.\ Kirchbach, Mod.\ Phys.\ Lett.\ 
{\bf A12}, 729 (1997), hep-ph/9701266.
\bibitem{SMW} X.\ Song, J.\ S.\ McCarthy, and H.\ J.\ Weber, Phys.\ Rev.\ 
{\bf D55}, 2624 (1997); T.\ P.\ Cheng and L.-F.\ Li, hep-ph/9701248; and 
refs. therein.
\bibitem{PAMD} P.\ A.\ M.\ Dirac, Rev.\ Mod.\ Phys.\ {\bf 21}, 392 (1949).
\bibitem{SK} L.\ Susskind, Phys.\ Rev.\ {\bf 169}, 1535 (1968); J.\ B.\ 
Kogut and D.\ E.\ Soper, ibid.\ {\bf D1}, 2910 (1970); S.\ Weinberg, ibid.\ 
{\bf 150}, 1313 (1966). 
\bibitem{LCQM} I.\ G.\ Aznauryan, A.\ S.\ Bagdasaryan and N.\ L.\ Ter-Isaakyan,
Sov. J.\ Nucl.\ Phys.\ {\bf 36}, 743 (1982).
\bibitem{W} H.\ J.\ Weber, Phys.\ Rev.\ {\bf D 49}, 3160 (1994).
\bibitem{WK} W.\ Konen and H.\ J.\ Weber, Phys.\ Rev.\ {\bf D41}, 2201 (1990);
Z.\ Dziembowski, Phys.\ Rev.\ {\bf D37}, 778 (1988).   
\bibitem{BS} S.\ J.\ Brodsky and F.\ Schlumpf, Phys.\ Lett.\ {\bf B329}, 111 
(1994).
\bibitem{pbar} W.\ Grein and P.\ Kroll, Nucl.\ Phys.\ {\bf A377}, 505 (1982). 
\bibitem{MAMI} L.\ Tiator, C.\ Bennhold, and S.\ Kamalov, Nucl.\ Phys.\ 
{\bf A580}, 455 (1994).
\bibitem{SW} M.\ J.\ Savage and J.\ Walden, hep-ph/9611210. 
\bibitem{KW} M.\ Kirchbach and H.\ J.\ Weber, nucl-th9612010, Comm.\ Nucl.\
Part.\ Phys.\ (1997), to be published. 
\bibitem{NA51} NA51 Collaboration, A.\ Baldit {\it{et al.}}, Phys.\ Lett.\ 
{\bf B332}, 244 (1994). 
\bibitem{KS} L.\ M.\ Sehgal, Phys.\ Rev.\ {\bf D 10}, 1663 (1974); G.\ Karl,  
ibid.\ {\bf D 45}, 247 (1992). 
\bibitem{S} M.\ Casu and L.\ M.\ Sehgal, Phys.\ Rev.\ {\bf D55}, 2644(1997). 
\bibitem{Lev} F.\ M.\ Lev, Nucl.\ Phys.\ {\bf A567}, 797 (1994).   
\bibitem{WW} J.\ Cohen and H.\ J.\ Weber, Phys.\ Lett.\ {\bf B165}, 229 (1985),
H.\ J.\ Weber and M.\ Weyrauch, Phys.\ Rev.\ {\bf C32}, 1342 (1985). 
\bibitem{PDG} Particle Data Group, Phys.\ Rev.\ {\bf D54}, 1 (1996).
\bibitem{AB} S.\ Adler and W.\ Bardeen, Phys.\ Rev.\ {\bf 182}, 1517 (1969);
G.\ Altarelli and G.\ C.\ Ross, Phys.\ Lett.\ {\bf B212}, 391 (1988).
\end{thebibliography}
\end{document}